# 水的结构和反常物性

姚闯[1]　张希[2,*]　黄勇力[3]　李蕾[1]　马增胜[3]　孙长庆[1,4,*]

1. 重庆市超常配位键工程与先进材料技术重点实验室，长江师范学院，重庆 408100
2. 深圳大学纳米表面科学与工程研究所，深圳 518060
3. 湘潭大学材料科学与工程学院，湘潭 411105
4. 南洋理工大学电气与电子工程学院，新加坡 639798

## 摘要

自从 Bernal-Fowler-Pauling 在 1933-1935 年间提出氢质子在两个氧原子之间的非对称等价位置以 THz 的频率自发往复隧穿后，液态水的结构尤其是水分子的近邻配位数目一直是学界关注的焦点。尽管水分子的刚性或柔性偶极子相互作用表述、纳晶非晶混相结构或均相涨落模型等假说已逐渐成为认知主流，但定量破解水在外场作用下所呈现的各种反常物性的进展依然缓慢。譬如，浮冰、复冰、超滑、热水速冻等现象的机理及内在关联仍有待系统深入研究。本文旨在尝试解读当前关注焦点和介绍最新研究进展的同时，融合连续介质论、分子时空论、质子量子论和氢键弛豫极化论并强调从传统的分子"偶极子-偶极子"到"氢键(O:H—O)超短程非对称强耦合"作用以及从源头的"质子隧穿失措"到"氢键受激协同弛豫"的思维转变。证据表明，水中键合质子数目和孤对电子数目和氢键的构型守恒和分子空间取向和质子隧穿规则应为关注焦点；通过氢键作用的静态四配位均相类单晶结构和动态强涨落可能是打破僵局的关键；由于氢键的 O:H 和 H—O 分段比热的差异，液态与固态之间存在一个具有冷胀热缩和相边界可调特性的准固态；键序降低导致氢键分段协同弛豫且使低配位水分子形成具有超低密度、强极化、高弹性、高热稳定性的超固态。由于 O:H 非键无处不在且起主导作用，拓展对于水溶液的认知到其它领域如含能材料的储能-燃爆机理、药物、食品、生命科学等会更加引人入胜，意义深远。

**关键词：** 氢键；温度；配位；压强；声子谱学

本项目得到科学挑战专题资助(TZ2016001;姚闯)，国家自然科学基金(No: 11502223；黄勇力)，湖南省自然科学基金(No: 2016JJ3119；黄勇力)和深圳市人才基金(No: 827000131；张希)的资助。

The work was supported by Science challenge project (TZ2016001; Yao Chuang), National Natural Science Foundation of China (No. 11502223 (Huang Yongli)), Hunan Natural Science Foundation of China (No. 2016JJ3119 (Huang Yongli)), Shenzhen Municipal Human Resources Fund (No. 827000131 (Zhang Xi))
*corresponding author e-mail: zh0005xi@szu.edu.cn ; ecqsun@gmail.com

# Structures and Anomalies of Water

Chuang Yao[1], Xi Zhang[2,*], Yongli Huang[3], Lei Li[1], Zengsheng Ma[3], Changqing Sun[1,4,*]

1. Key Laboratory of Extraordinary Coordination Bond Engineering and Advanced Materials Technology (EBEAM) Chongqing Municipality, Yangtze Normal University, Chongqing 408100
2. School of Mechanical and control Engineering, Shenzhen University, Shenzhen 518060
3. School of Materials Science and Engineering, Xiangtan University, Xiangtan 411105
4. Nanyang Technological University, Singapore 639798

## Abstract

The structure of liquid water particularly the number of bonds per water molecule has been a debating issue since 1933-1935 when Bernal, Fowler, and Pauling firstly proposed the scenario of proton "transitional quantum tunneling" in THz frequency at asymmetrical sites between two oxygen ions. Although conventions of the rigid or flexible dipole-dipole interaction, nanophase mixed amorphous structure or homogeneous fluctuating phase models, solute diffusion dynamics or hydration length scale premises have been becoming dominant, mysteries such as floating of ice, regelation of ice (compression melting), slipperiness of ice, fast cooling of warm water, etc., have yet to be resolved. The definition of hydrogen bond needs yet to be certain. In this perspective, we emphasize that it would be more efficient to transit the conventional "dipole-dipole" interaction to "hydrogen bond (O:H—O) asymmetrical, short-range, correlative" interaction, from the "proton translational tunneling" to "hydrogen bond cooperative relaxation". Progress revealed also that the O:H—O bond configuration and the numbers of protons and nonbonding electron lone pairs conserve and that water forms the tetrahedrally-coordinated, strongly correlated, fluctuating single liquid crystal. The O:H nonbond and the H—O bond segmental specific heat disparity derives a quasisolid phase between the liquid and the solid. With tunable boundaries, the quasisolid phase possesses the negative thermal expansion coefficient. Remarkably, molecular undercoordination results in a supersolid phase that is highly polarized, thermally stable, viscoelastic, and lesser dense. Extending hydrogen-bond knowledge to the energy storage – explosion reaction mechanics of energetic materials may further verify the comprehensiveness and universality of the current notion of hydrogen bond cooperativity – nonbonding interaction is ubiquitously important.

# Table of Contents

## 1. 引言：机遇与挑战

水是生命的起源 – 如果没有水，生命即不能维持也不能繁衍[1]。然而，正如《Nature》杂志前资深编辑菲利普·鲍尔所言，水的确是太神奇、太反常、太具有挑战性了，甚至于说不可能真正有人知道水到底是什么[2]。它多变的相结构和反常物理性能吸引了无数的智慧头脑试图厘清其特殊性本质原因和各种物性的内在关联。《Science》杂志2005年将水的结构列为人类所面临的125个难题之一[3]。美国化学学会出版的《Chemical Reviews》在2016年6月集结了业界权威发表共十余篇专辑文章从方方面面介绍了液态水的研究现状和进展[4-12]。《Reviews of Modern Physics》2016年2月发表的一篇文章指出[13]，人们对水的研究最多，但知之最少；研究投入越多，手段越先进，认知越迷茫[14]。从表1列出的当前最具代表性的关注焦点和主流认知实践栏目可见，辩论仍在继续，对冰、水和溶液的各种反常物性的共同机理达成共识，尚需时日。

表1. 关于水专题的关注焦点、主流认知实践和研究新探
Table 1. Focuses, debates and recent explorations

| 关注焦点 | 传统认知 [4-14] | 悬疑新解 [1, 15-17] |
|---|---|---|
| 配位规则 | 或非四配位动态结构 | 质子和孤对电子数目以及氢键(O:H—O)构型守恒 |
| 作用形式 | 刚性或柔性偶极子近似 | O:H—O 非对称超短程耦合振子对 |
| 势能函数 | 双体或多体、对称或非对称 | 三体非对称、超短程、强耦合 |
| 输运规则 | $2H_2O \leftrightarrow H_3O^+ + HO^-$ 自发质子隧穿 | O:H—O 分段长度和能量受激协同弛豫 |
| 液态结构 | 均相涨落或高低密度混相 | 超固态表皮包裹的强涨落单晶 |
| 温致特性 | 经典热力学；混相权重调控 | O:H—O 分段比热叠加；液态-准固态-固态温区密度振荡 |
| 配位弛豫 | 过热和过冷；力热稳定性 | 超固态：准固态相边界受激色散伴随电子极化密度致疏 |
| 压致特性 | 质子平移隧穿和 O—O 收缩 | O:H 缩短 H—O 伸长伴随极化 |
| 冰皮润滑 | 准液态(压融,热摩擦)润滑 | 界面超固态弱声子高弹性和极化电子相斥 |
| 压融复冰 | 表皮液态联接剂 | O:H—O 变形破损自愈合；H—O 键能主导熔点 |
| 温水速冻 | 多因素定性猜测 | 能量存储-释放-传导-耗散：氢键记忆与表皮超固态 |
| 盐水溶液 | 溶质迁移和水合层厚 | O:H—O 网络离子局域极化调制氢键刚度、序度和丰度 |
| 酸合溶液 | $H[H_2O_2]^+$ 质子隧穿扩散迁移 | $H_3O^+$ 和 H↔H 反氢键点致脆退极化 |
| 碱合溶液 | $HO^-$ 迁移 | $HO^-$ 和 O:⇔:O 超氢键点压缩极化 |
| 电致相变 | 阿姆斯壮水桥 | 极化拓展准固态相边界 |
| 液态抗磁 | 青蛙磁悬浮 | 感生偶极子电流对抗源磁场 |

Table 1. Focuses, convention and recent resolution

| Focuses | Convention [4-14] | Resolution [1, 15, 17, 18] |
|---|---|---|
| Coordination Rule | Dynamically ulceration | 2N number and HB (O:H—O) configuration conservation |
| Interactions | Rigid or flexible dipoles | HB asymmetrical oscillating pair |
| Potentials | Two- or three-body, symmetric or asymmetric | Three-body, asymmetric, ultra-short-range, strong correlated |
| Transport dynamics | $2H_2O \leftrightarrow H_3O^+$:$HO^-$ proton tunneling | HB segmental cooperative relaxation; tunneling forbidden |
| Structure order | Amorphous, uniform, or mixed HDA and LDA | Supersolid skin covered fluctuating single crystal; regulated molecular orientation |
| Thermal excitation | Classical thermodynamics; modulated mixed phases | HB segmental specific heat superposition, four-zone density oscillation |
| Undercoordination | Supercooling and superheating | Supersolidity: quasisolid phase boundary dispersion and electron polarization, density loss |
| Compression | Proton centralization by translating tunneling | O:H contraction and H—O extension accompanied by polarization |
| Ice slipperiness | Quasiliquid layer lubricant (pressure melting, friction) | Skin supersolidity, soft phonon high elasticity, polarization repulsion |
| Ice regelation | Liquid adhesion | HB self-recovery, H—O energy dictates $T_m$ |
| Hot water cools faster | Multi-factor speculation | Energy storage-emission-transfer-dissipation: HB memory and skin supersolidity |
| Salt solvation | Water structure maker or breaker | Ionic polarization transits HB stiffness, order and abundance |
| Acid solvation | Proton tunneling diffusion; cluster formation | H↔H anti-HB fragilization and depolarization |
| Base solvation | $HO^-$ diffusion | O:⇔:O super-HB compression and polarization |
| Electric polarization | Armstrong water bridge | quasisolid phase boundary dispersion |
| Anti-magnetism | Frog magnetic suspension | Moving dipoles under Lorentz force |

"液态水的结构"是一个典型的科学难题。其主要原因是定量的理论计算与实验测量结果不能完全匹配；常规液体-固体相变理论难以对水进行合理描述而且不同的实验方法，如 X 射线和中子散射，以及红外与拉曼光谱和动力学测量等方法得到的实验结果也很难自洽。所以确定液态水中的动态氢键结构是一项极具挑战性的任务[13]。例如，《Science》2004 年发表了一篇文章指出液态水中每一个水分子主要是以略多于二配位的链状氢键结构为主，而不是惯常的四配位结构[19]。这篇文章当时引起了一场很大的轰动并被《Science》评为当年的十大科学进展之一。不过编辑在推荐此项工作时谨慎地指出，先不必急于修改教科书，因为争论还会继续。果然，Smith 等[20]同年在《Science》杂志发文支持水分子的四配位结构。Head-Gordon 等[21]2006 年在《PNAS》上也发文指出，采用水的四配位结构同样能够拟合[19]的实验结果。《Physical Review Letters》在 2008 年发表了 Hermann 等[22]的计算光谱结果认为传统的四配位结构更为合理。最近北京大学江颖等通过扫描隧道显微技术观察直接证明了氧的 sp³-电子轨道杂化结构和单个水分子

在极低温度、超高真空和超低配位下的氢质子的核量子效应[23, 24]。

此外，人们通常把水分子类比成刚性或柔性偶极子以研究它们在偶极子海洋中的相互作用及行为，譬如 TIPnQ(n = 1-5) 系列模型[1]。此类模型把单个水分子视为一个偶极子并用 n 个相对固定位置的固定电荷表述，而专注考虑这些偶极子间的相互作用。这些模型忽略了非常关键的分子内和分子间的强关联耦合作用。微观上，普遍认为水是纳米晶体、玻璃体、或者是玻璃体镶嵌的纳米晶体具有不同密度的混相结构[13]。所以认为高低密度两相的体积比随温度的改变是浮冰现象的根源。虽然大多数实验和理论的研究论文都声称支持原来的四配位结构，但依然难于从理论上一致地定量地描述液态水的结构与其反常特性的关联。

从某一特定角度出发，各个模型都有道理而且表述正确，但难免有其局限性。关于水和冰的研究现状可以归纳为一种现象伴随着多种互不相让的理论辩争，而急需一个普适的理论把水和冰的结构和所有的反常物性统一起来，以揭示其本质和规律。只有在水的基本构序规则和氢键受激弛豫动力学图像清晰之后，我们才有可能探索水和溶液的反常物性和酸碱盐的水合动力学以及对溶剂的氢键网络和物理性能的调制。

尽管人们已经做出了很多努力，水科学研究进展仍远不及人意。关于冰水的研究现状可以归纳为一种现象伴随着多种互不妥协的理论争辩。此领域更是学派林立，学说专深。就某一专题达成共识，似乎尚需时日。人们梦寐以求的是如何用一个系统的理论统一冰水的所有的反常物性，以揭示其本质和运行规律。水科学研究进展缓慢的主要原因有以下几个：

1) **神秘不可认知论**。菲利普·鲍尔 2014 年在《欧洲物理杂志》关于水研究的特刊上指出[25]：虽然承认这一点的确很尴尬，但是没有也不可能有人真正懂得水。更糟糕的是，对水的研究投入最多，知之甚少。投入越多，问题也越多：新的技术更深层次探索着的液态水结构，也引出了更多的谜题[13]。因此，人们总是预设前提，认为太复杂太难了。也有很多人认为水是神的使者，只有虔诚地祈求上帝告知它的秘密。这难免使人望而生畏。

2) **唯权威或传统是从论**。避开 Nature 或 Science 发文或某权威理论的约束，我们应该独立思考细究、批判吸收、去伪存真以寻找科学规律和可控因素。任何理论或学说都有时间的局限性。我们要尊重但不唯权威或传统。譬如，博奈尔-福勒-泡令（1933）以及葛洛哈斯（Grotthuss, 1906）在由泡令 1935 年首次提出[26]氧的 $sp^3$ 电子轨道杂化论之前提出的水中[27, 28]和酸溶液中[29, 30]的质子随机隧穿机制。因当时认知的局限而只能从表观 O⋯H⋅⋅O 非对称库仑作用而无法从 $H^+$ 与其左右两侧氧离子的结合能的差异[15, 31]的角度判定隧穿是否可行。

3) **以偏概全，重表观轻本质和起因**。在很多情形下，人们往往局限于采用某一特定方法处理特定条件下特定变量的精确求解，而忽略所有涉及的变量的关联。譬如，超微水滴在真

空中和高空稀薄大气层的云雾的结冰，既有相对大气负压强，又有低温和分子低配位的同时介入，这三个变量的耦合导致表观的过冷和过热现象。理解奇异现象背后的本质以及决定本质的起因具有更大的挑战和乐趣。

4) **液相非晶或多相结构不确定论**。人们往往从直觉上认为液态即是非晶，至多是多晶。结构的失序和不确定妨碍对耦合作用的正确表述。将液态系统分解成稳态结构和动态涨落的叠加，而稳态作用势的正确表述是关键。只能通过实验测量和数值转换获得包含外场激励的作用势，从而诊断液态水的晶体和物相结构。

5) **分子电子运行规则的认知局限性**。O:H—O 耦合氢键与分子内 H—O 强的共价键、分子间 O:H 弱的非键和 O··H··O 近等非对称库伦作用混淆。在氧的 $sp^3$ 电子轨道杂化被发现之前的 O··H··O 库伦非对称作用认知局限是可以理解和接受的。然而，质子隧穿的局限已经束缚了研究的进展而水分子间与分子内的作用及其耦合的恰当表述至关重要。作为水的基本结构和储能单元，O:H—O 键决定分子的堆垛形式和时空行为以及一切与尺度和能量有关的物理量。水中质子和孤对电子的数目及行为必须遵守相应的规则。

对冰水研究可以归纳为以下几种可以互补且卓有成效的多尺度研究方法：

1) **经典连续介质论** [32-35]：由单质固体和理性气体的统计热力学衍生而来的，这种方法是从焓、熵、自由能的角度成功地处理许多宏观物理量诸如介电性、表面应力、粘滞性、扩散系数、液/汽相边界的描述。这种方法将外界激励作为可测宏观物理量的自变量而它的局限性而忽略分子间的弱作用甚至是溶液中分子间的排斥作用。

2) **分子时空论** [36-40]：这种方法把水分作为独立的可伸缩或刚性偶极子处理。分子动力学计算与超快光谱结合研究以分子作为基本单元的时空行为。获得的信息包括分子在某一位置的滞留时间、扩散系数、溶液粘滞度等。它的局限性在于如何考察分子间与分子内作用的耦合以及外场激励的介入方式。

3) **质子的量子论** [41-43]：与超低温、高真空、超低配位条件下的扫描隧道显微术结合，从头开始路径积分 (PIMD) 已经观察到质子的集体隧穿行为以及确定的质子的和量子效应对 H—O 零点真动能的贡献。从实验上证明至少在 5 K 温度下，氧的 sp 轨道杂化发生以及单个水分子保持四配位构型。质子的量子作用使 O:H—O 键的较长分段加长而使短段缩短。

4) O:H—O 氢键协同弛豫与非键电子极化论 [1, 15, 44, 45]：这一理论关注 O:H—O 键的分段强度和比热的非对称以及耦合作用导致的分段受激协同弛豫和电子极化。将拉格朗日真动力学、分子动力学、密度泛函理论、以及电子声子计量谱学技术相结合不仅探测到氢键的受激协同弛豫和水合转换动力学和热力学而且可以把声子的丰度—刚度—寿命—序度的受激协同转换与液体的粘滞性、表面应力、相边界调制以及想变得临界温度和压强等。

融合这些经典介质论、分子时空论、质子量子论、氢键弛豫极化论，将逻辑分析、理论计算、实验测量相结合才可能有所突破。

## 2. 对策：思维的转变

当我们对某一主题付出长期和大量的努力而进展不如愿时，我们有必要重新审视所涉及的基础理论或假设是否健全或有所疏忽。如果寻求对液态水的结构及其反常物性理解的突破，我们应该从源头重新思考[1, 15]。泡令[26]1939 年提出，化学键的属性是连接物质的结构和属性的桥梁。这是我们试图改变思维方式的动力和基础。要想摆脱目前的困境，应该从水中氢键(O:H—O)的形成和受激弛豫以及相应的电子行为着手建立各可测物性之间的关联[46]。相应地，我们采用一条氢键作为最基本的结构和储能单元，并专注于这条代表键和电子的受激弛豫行为以及它的演变对可测物理量的贡献。实践证明，更换角度思考和探究，或许可以得到意想不到而且自洽的结果。

在尝试过程中，我们强调从传统的水"偶极子-偶极子"间的相互作用的表述过渡到"O:H—O 键超短程非对称强耦合"作用，从源头的"质子隧穿失措"转变到"氢键受激协同弛豫"的思维方式[1]。证据表明[1, 15]，水中键合氢质子和孤对电子的数目以及氢键的构型应为起始关注焦点；通过氢键作用的静态四配位类单晶结构和强涨落是打破僵局的关键；由于氢键分段比热的差异，液态与固态之间存在具有冷胀热缩和相变解可调特性的准固态；由于键序降低导致的氢键分段协同弛豫，低配位水分子形成具有超低密度、强极化、高弹性、高热稳定性的超固态；O:H—O 的压致对称使准固态相边界拓展而导致复冰现象。作为新的尝试，我们拓展冰水表面的概念到有一定厚度的冰水表皮或表层；作为当前普遍采用的和频介电谱（表层偶极矩取向）和超快红外谱（声子寿命）技术的补充，我们拓展传统的稳态红外吸收和拉曼散射谱学测量实验到计量差谱分析以直接获得关于氢键分段刚度（特征频移）、序度（半高峰宽）和声子丰度（峰面积）受激变化信息。这一系列的思维转变和相应的概念方法确实能够使我们系统地检测氢键的 O:H 和 H—O 分段长度和能量的协同弛豫以及非键电子极化对水和冰的宏观可测物性的主导作用。

## 3. 原理：禁戒与守恒
### 3.1. 质子—孤对数目与氢键的构型守恒

早期关于氧吸附研究揭示[46, 47]，在与其它任意电负性较低的原子发生化学反应过程中，一个氧原

子首先从两个不同的近邻异质原子各俘获一个电子后杂化其自身的 2s2p 电子轨道而与近邻氧原子结合而形成二重对称的准四面体结构。每个氧原子含有两个键合质子（H+）和两个孤对电子（:）而形成水分子的四配位构型。无论是在汽态，液态，还是固态，甚至是 5 开尔文温度下四面体构型的水分子都保持稳定[48-50]。如果所考虑的水样本中包含 N 个氧原子，那末这个样本中就分别恒有 2N 个与氧键合的质子和孤对电子存在。这 2N 个质子和 2N 个孤对电子结合决定了 O:H—O 键的构型并作为水和冰的基本结构和储能单元。根据键弛豫理论[46]，氢键的受激弛豫及其电子行为主导水的结构演变及其所呈现的物性。采用质子和孤对电子的 2N 数目和氢键构型的概念要比讨论水分子的配位数目或单个水分子偶极子更加意义深远。

氢键具有很强的柔性和可极化特性而且普遍存在于水和冰的各相中，它也是构成水溶液的氢键网络的主体。氢键的关键是非键孤对电子的存在。即使在 $OH^{3+}:OH^-$ 超离子态（存在于 2000 开尔文和 $2\times10^{12}$ 帕超高温高压条件下[51]）和分段长度对等的第 X 相中（存在于 $6\times10^{10}$ 帕超高压和任何温度[52]），氧-氧之间的键合性质和构型不变，只有氢键的键角、取向、分段长度和刚度可以发生弛豫和涨落。氢键的受激弛豫唯一地决定水的各相结构和在受激时展现的各种性能[53-56]。

相对传统的"偶极子—偶极子"相互作用的处理方法，氢键非对称耦合振子对的分段协同弛豫和非键电子极化具有更普遍的意义，并且可通用于所有包含孤对电子的，如氨($NH_3$)和氟化氢($HF$)，甚至是三硝基甲苯（$C_7H_5N_3O_6$）炸药等系统中。此外碳氮氧氟等原子在化学反应过程中，它们的 2s2p 电子轨道也发生杂化与客体原子形成四面体结构单元，但各自所含的质子和孤对电子的数目有所不同。

## 3.2. 分子取向规则与质子隧穿能量禁戒

由于 2N 和 O:H—O 键的构型守恒，水分子的空间取向必须保持在一定的范围内转动。而单个分子发生大角度转动应当禁戒以避免 O:H—O 失配产生 H↔H 和 O:⇔:O 排斥。当转动角度大于 60°时，O:H—O 失配发生而水的结构失稳。如图 1 所示，每个水分子周围的四个分子中必须有两个以 H+ 而另外两个以":"指向中心分子的":"或者 H+ 以形成稳定的 O:H—O 键。即使额外的 H+ 和":"分别以 $H_3O^+$ 和 $HO^-$ 四面体替位形式介入，周围是分子的空间取向也不会改变。

当 HX 酸、YOH 碱、和 YX 盐水合后，溶液中即产生额外的 H+、":"、X- 和 Y+ 离子(X = Cl, Br, I; Y = Li, Na, K)。$H_3O^+$ 和 $HO^-$ 分别有三个质子和三个孤对电子且保持其四配位构型。$H_3O^+$ 的介入产生一对未能与":"配对的质子并形成 H↔H 排斥，而破坏纯水的 2N 守恒。由于 H↔H 近邻质子间的排斥，我们称之为反氢键点断裂元以为它破坏氢键网络结构和稀释液体。同理，$HO^-$ 的介入产生 O:⇔:O 超氢键点压缩元并且具有宏观压力同样或者更强的功能。作为点电荷，酸碱盐中的 X- 和 Y+ 离子的各自径向分布电场极化其周围的水分子而形成强极化水合层，如图1所示。酸碱盐水合并不改变水分子间的相对取向， 而只是由于排斥和极化作用改变氢键分段的能量和长度。

对于氢键的分段弛豫和非键电子的极化，酸水合与加热效果相同，碱水合与加压效果相同，盐水合与分子低配位效果相同，虽然微观机理不同[56]。

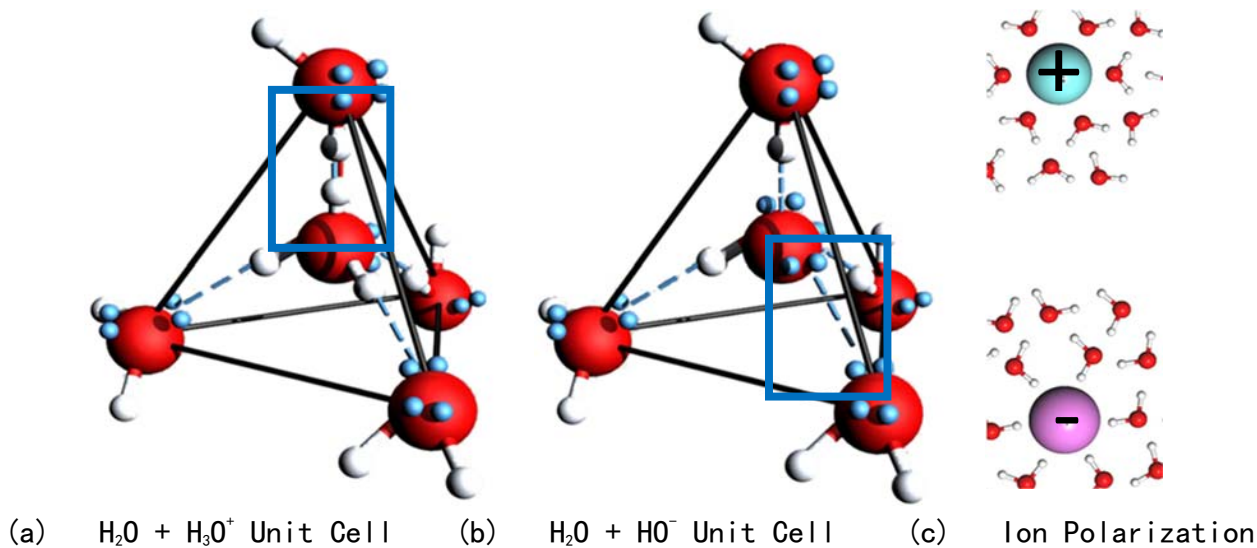

(a)　　H$_2$O + H$_3$O$^+$ Unit Cell　　　(b)　　H$_2$O + HO$^-$ Unit Cell　　　(c)　　Ion Polarization

图1. 水分子空间取向规则以及2H$_2$O 元胞中心(a) 被酸溶液[57]的H$_3$O$^+$ 和(b)碱溶液[58]的四配位HO$^-$替代后的元胞结构 和(c) 离子电场对水合层内分子的极化取向规则[59]。
Fig. 1. Orientation regulation for water molecules. Solvation of (a) acid [57], (b) base[58] creates excessive H$^+$ and ":" and create the 4-coordinated H$_3$O$^+$ and HO$^-$ without changing the orientation of its surrounding water molecules. (c) Ions in solutions serve each as source of radial electric field that clusters, aligns, and stretches and polarizes the O:H-O bond in the hydration shells without changing the relative orientation of other water molecules [59].

此外，O:H—O 键分段能量的巨大差异（0.1 – 4.0 电子伏特）禁止质子与其近邻":"易位而在两近邻氧原子间发生平移隧穿。由于集体转动造成表观跃迁隧穿应当别论。氢质子在两个氧之间发生的 "质子平移隧穿失措"[27, 28]是指氢质子在两个氧原子间等价位置上以太赫兹的频率等几率自发地往复运动，也即发生2H$_2$O ↔ H$_3$O + HO 的超离子态转变。事实上，H—O 结合很强，至少需要4-5电子伏特的能量或吸收121.6纳米波长的激光才能使其断裂[31]。计算研究表明[51]，2H$_2$O ↔ H$_3$O + HO 的超离子态转变只有在极端2000K 和2TPa压强下才能发生。

水的 2N 质子和孤对电子数目以及氢键构型守恒与分子空间转动和质子平移隧穿禁戒规则决定了水的静态单晶和动态强涨落特征。确切地讲，水是由低配位分子组成的超固态表皮包裹的四配位具有核-壳结构双相单晶，既不是非晶也不是多晶。氢键的 H—O 分段能量和长度的受激弛豫决定水的能量吸收和释放，而 O:H 的弛豫主导序度、涨落、扩散、甚至汽化及其能量耗散。

## 3.3. 氢键作用势与水的强涨落单晶结构

首先，关于氢键的定义需要统一。有人指分子间的 O:H 非键或分子内的 H—O 共价键为氢键，水分子间的由孤对电子构成的 O:H 非键和水分子内的极性共价键 H—O 的组合，而不是它们的任一分段[1]。作为振子对，氢键包含超短程，非对称的 O:H 和 H—O，并通过近邻氧离子上的电子对之间 O—O 的排斥而耦合。正是这些往往被忽略的氢键的非对称性，超短程作用和氧-氧间的强关联耦合主导水和冰在面对任何程度的微扰和辐射时所显示的超常的自适应性、协同性、自愈合和高敏感等特性[1]。氢键的非对称（分别为~0.1 和~4.0 电子伏特的结合能）和氧-氧的强关联使水分子作为一个整体在一定的范围内不停的振动和集体转动以逼近其理想的统计平均单晶结构.

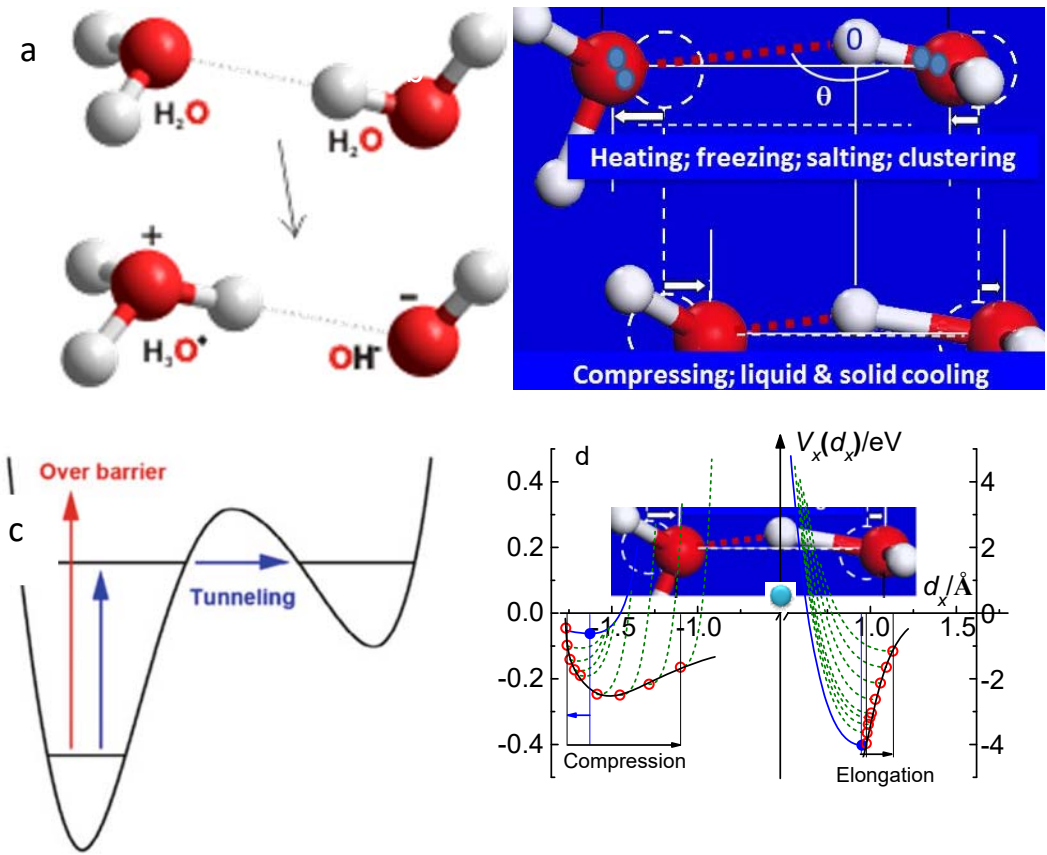

图 2 (a)传统的"质子隧穿失措"模型[27, 28]和(b)氢键受激协同弛豫[15]以及相应的(c)质子隧穿[60](d)和氢键受压[54]作用势函数。以氢质子为坐标原点，两个氧原子在非保守力作用下沿着氢键朝相同方向以不同的程度位移。(d)中的两侧的蓝点对应不计及氧-氧排斥时各段势能的平衡点；近邻的红点对应氧-氧排斥介入后与各段势能之和的平衡点；其余的从左到右的红点依次对应在非保守力（压强从大气环境增加到 6×10¹⁰ 帕）作用下总势能的平衡点。

Fig. 2 (a) The "proton translational tunneling" model for the 2H$_2$O ⇔ H$_3$O$^-$ + HO$^+$ random transition [27, 28] and (b) the "hydrogen bond cooperativity" mechanism for the O:H—O bond relaxation under mechanical compression. The corresponding potential model (c) for (a) [60] and for (d) [54]. Under mechanical compression and O—O Coulomb repulsion, both oxygen ions dislocate along the O:H—O bond in the same direction by different amounts with respect to the $H^+$ coordination origin. All symbols correspond to equilibrium of the interatomic potentials with (red) and without (blue) O—O repulsion involvement. From left to right is the potential paths of the O:H—O bond under mechanical compression from the ambient to 60 GPa.

图 2 比较(a)Bernal-Fowler-Pauling 的"质子隧穿失措"模型[27, 28]所描述的 $2H_2O \leftrightarrow H_3O:HO$ 超离子态转变[51]和(b)氢键受激弛豫模型[15]以及(c, d)相应的作用势函数[44, 60]。将实验测得的氢键的 O:H 和 H—O 分段长度和振动频率的受激弛豫通过拉格朗日转换成各自的力常数和结合能而得到图 1d 所示的氢键受压弛豫势能路径[54]。

3.4. 氢键受激协同弛豫与水的反常物性

图 2b 所示氢键的受激弛豫动力学。在外加非保守力场驱动下，氢键的强弱两段永远以主-从方式协同弛豫。如果其中一段伸长，那么另一段就收缩。所以，选取看似不停运动的氢质子为坐标原点，它的两侧的氧离子总是以不同的幅度沿着连线朝着相同方向位移。因为结合强度的不对称，O:H 非键的位移量总是大于 H—O 共价键的位移量。氧-氧间距的任何变化都是通过氢键两分段的一伸一缩来实现的。此外，任何一段的伸长和收缩都伴随其刚度的减弱和增强而且可以用红外或拉曼声子谱学方法直接观测。振动谱的频移与相应分段的刚度的方均根相关。键的刚度是其杨氏模量与其长度的乘积或者其能量与长度平方的商。拉曼特征峰蓝移代表相应分段长度缩短，结合能和刚度增强，反之亦然[15]。此外，氢键的分段弛豫总是伴随着非键电子极化或退极化。

实验结果显示，O:H 段在 4 ℃ 左右为 1.7 埃长并具有 0.1 电子伏特的结合能。其刚度决定<200 $cm^{-1}$ 波数范围内的特征声子振动频率以及结冰温度、水的德拜温度、以及沸点和露点温度[15]。H—O 段为 1.0 埃长和 4.0 电子伏特的能量以及 3200-3600 $cm^{-1}$ 波数范围的特征声子振动频率。H—O 段的刚度决定冰的融化温度和氧的 1s 能级移动等。非键电子的极化增强水溶液的粘滞性、弹性、表皮应力、疏水和润滑性能等。这里所指的非保守力场包括温度、压强、低配位驰豫势场、电场、磁场、水合溶质的离子电场等。

4. 方法：谱学与解析

研究水的谱学方法很多[61]，包括中子散射，X-射线散射和带边吸收或发射，光电子发射等[5, 9, 11]。其中红外-可见光和频谱(SFG)技术测定表界面的介电性能或分子偶极矩的取向。时间分辨瞬

态红外吸收谱(t-2DIF)探测声子的弛豫时间或寿命以获取液体中溶质和溶剂分子的动力学信息并以此标定溶液的粘滞性。作为上述谱学方法的补充，我们采用拉曼散射差谱分析技术（DPS）探测具有分子空间位置分辨功能的有关氢键的O:H和H—O分段协同弛豫动力学行为[1, 18]。声子谱的某个拉伸模式的特征峰代表在实空间中具有相同振动频率的所有的键的傅立叶变换，与这些键的空间取向和所在位置无关，与其所在的相结构无关。峰的形状是分布函数，而它在特定频率区间的积分是声子丰度，对应参与所能测量振动的键的数目。这样我们可以探测水及其溶液在外场驱动下氢键的分段长度和刚度变化，局域分子涨落序度，和声子丰度的弛豫和转变。受激发后谱峰面积、频率、和半高宽的变化对应氢键的数目、刚度、和序度的受激跃迁转换。通过前后测量的差谱分析，我们可以翔实地了解在不同的空间位置和配位环境下以及在外场作用下水溶液中的氢键网络到底发生了什么和怎样发生的。

此外，分析液体微射流的紫外和X射线光电子谱可以获得不同空间位置的氧原子1s能级的移动和非键电子的极化。求解拉格朗日方程可以有效地处理氢键耦合振子对，并将测得的氢键受激分段长度和振动频率弛豫转换成相应分段的力常数和结合能，继而得出氢键作用势在非保守外场驱动下的弛豫路径。傅立叶流体热传导方程的有限元方法的成功求解证明了氢键的记忆特性，水表皮的超固态，以及非绝热耗散是决定姆潘巴效应的三要素。数值解析计算和谱学测量分析相辅相成是验证逻辑推理的必要手段，而合适的理论模型是确保认知突破和自洽的核心。

在处理水这类高有序、强关联、强涨落系统时，我们应该关注所有相关参量集合的统计平均，而不是苛求某一特定参量在某一特定位置、特定条件下的即时准确性和精度。我们也应该更多关注具有明显意义、反映本质的物理量而把那些具有共性的犹如长程作用和非线性效应等高阶贡献暂时作为平均背景舍弃而直指问题的本质。把问题简化到不可再简化为止，去求所要解决的问题的低阶解，然后再把舍去的高阶参量逐一索回。尽管实验条件苛刻而且理论计算艰辛，结果会更加奇妙多彩。当考虑各阶近似的完整贡献后，我们可以得到对所有相关问题的有明确物理意义的完备通解及其关联。

5. 进展：谜题的探解
5.1. 温驱密度震荡：冷膨胀准固态与浮冰机理

浮冰是一个典型的历史谜题。最先由伽利略与路德维克·哥伦布于1611年在意大利弗洛伦萨展开了长达三天的激烈辩论[62]。他们从浮力定律，表面张力，和质量密度等不同角度进行争辩但没有得到明确的结论。为纪念关于浮冰辩论400周年，25名智者于2013年重聚弗洛伦萨用了一个星期时间专门讨论水的未解之谜。然而激烈的争论延续至今尚且不休,仍无定论。

由于传统相变理论在处理冰水热力学行为时受阻，我们有必要诉求新的思路。首先引入单键比热 $\eta x$ 的概念。因为氢键是由两段构成的，所以有必要引入单键比热的概念并考虑氢键分段的各自

比热的差异。两者的叠加才是水的比热。正是氢键分段比热的差异从本质上区分了水与其它可以用单键平均表述的物质的热力学行为。这也可能是传统热力学相变理论对冰水热力学失效的原因。固体比热的德拜近似有两个特征量。一是德拜温度决定比热曲线的饱和温度，二是比热曲线对温度的积分正比于相应分段的结合能。此外各分段比热的德拜温度正比于其特征拉伸振动频率，也就是服从爱因斯坦关系：$\Theta_{Dx} \propto \omega_x$。这样，氢键的分段的振频变化直接联系到氢键的比热曲线和冰水的热力学行为。

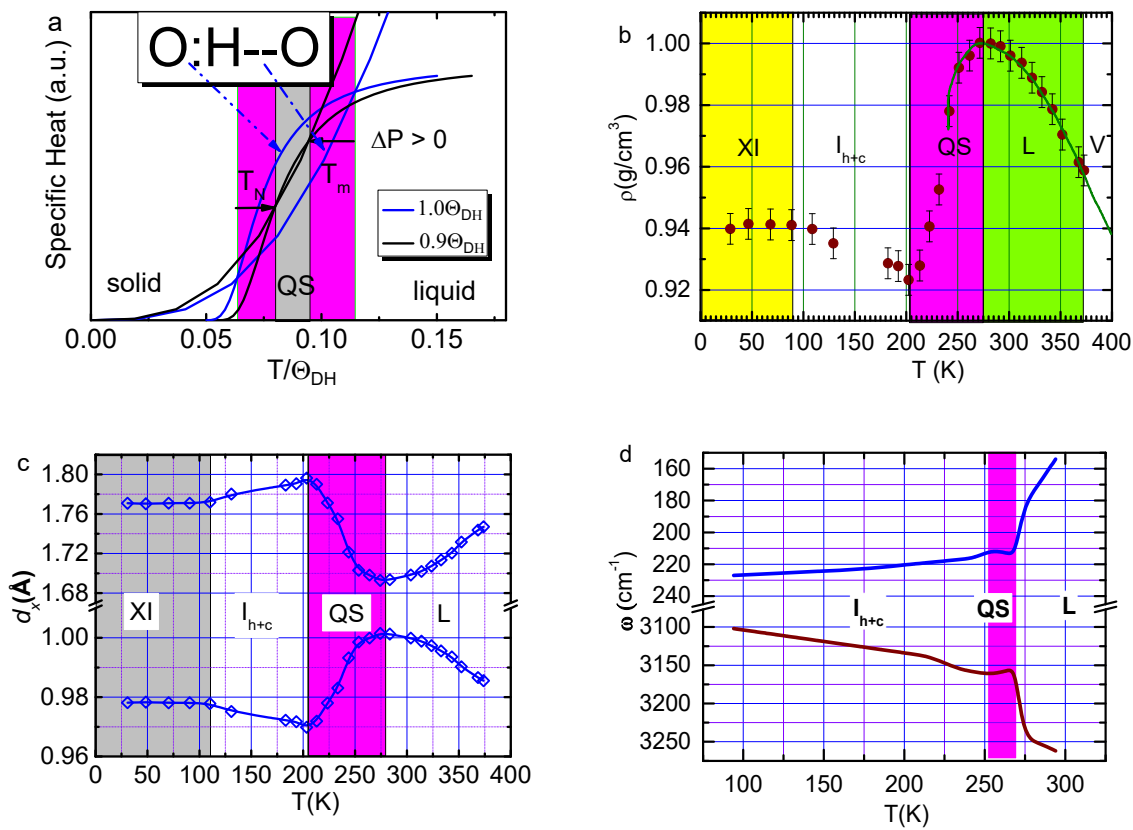

图 3 水的(a)氢键分段德拜比热模型，(b)实测质量密度[63]，以及由(b)转换的(c)氢键分段长度和(d)实验测量的特征拉曼频率随温度的变化[55]一致地显示这些物理量的四温区振荡效应。(b, c)中温度低于 273 K 对应 1.4 纳米水滴[63]而(d)对应大体积水。准固态 QS 相边界由 O:H 和 H—O 两个分段的振动频率决定。准固态反常热膨胀是浮冰的物理起因。

Fig. 3 (a) O:H—O bond segmental specific heat in Debye approximation, (b) measured temperature dependence of mass density, and derived temperature dependence of (c) the O:H—O bond segmental lengths and (d) vibration frequencies [55] show consistently oscillation of these properties in four regimes. Plots (b, c) lower than 273 K correspond to 1.4 nm sized droplet [63] and (d) the to the bulk water.

图 3 说明氢键的分段比热,冰水质量密度,以及氢键分段长度和振动频率随温度的变化的一致性[55]。氢键分段比热曲线的叠加产生两个交点并把整个温区分成五个具有不同比热比值$\eta_L/\eta_H$的温段:汽态 V($\eta_L = 0$),液态 L($\eta_L/\eta_H < 0$),准固态(quasisolid, QS) ($\eta_L/\eta_H > 0$), 固态 $I_h$ + $I_c$($\eta_L/\eta_H < 0$)和低温 XI 相($\eta_L \cong \eta_H \cong 0$)。$\eta_L$代表 O:H 的比热而$\eta_H$为 H—O 的比热。这两个交点分别对应密度的极大和极小值,对于纯水它们分别在 4°C 和 −15°C [55]。这两个交点分别靠近熔点和成核点。从液态到固态的转变需要跨过以这两个极值点为边界的准固态或称为准液态的温区。

氢键分段比热的比值$\eta_L/\eta_H$决定氢键的协同热弛豫行为。比热值较低的分段服从常规的热胀冷缩定律;由于氧-氧排斥,而另一段则冷胀热缩。由于 O:H 热弛豫量总是大于 H—O 段的弛豫量和 O—O 间距决定水的体积变化,液态水和 $I_{h+c}$固相冰在宏观上显示不同速率的热胀冷缩行为,而冷膨胀主导准固态的行为。因为在低温区,两支德拜比热曲线趋于零,所以固态第 XI 相的密度基本恒定,也就是氢键的长度和能量在此温区对温度不敏感。

更重要的是准固态的相边界可以通过氢键分段的刚度受激弛豫直接调制(图 3a),因为特征频率直接与德拜温度关联。例如加压缩小准固态温区而提升冰点并压缩熔点而导致复冰现象(压致熔点降低, 冰点升高)[64-66];水分子配位数降低可以扩大准固态温区而提升熔点并压缩冰点而导致单层水在室温下所显示的类固态行为和纳米微化时的过冷和过热现象[45, 67],以及纳米气泡的超常力学强度和热学稳定性[68]。在准固态温区 H—O 的冷缩和 O:H 的膨胀增大 O—O 间距而降低密度。所以浮冰现象发生。这一机理对其它反常冷膨胀物质如石墨和钨酸锆等可能具有参考意义 – 非对称超短程强耦合作用是反常冷膨胀现象的物理基础[1]。

## 5.2. 低配位超固态:冰皮润滑与水表皮超疏水

冰的表皮具有自然材料中最低的摩擦系数。复冰即是冰在受压时熔点降低但压力撤除后熔点还原。自从 1850 年开始,Faraday 等从液态表皮、受压熔化、和摩擦生热的角度试图解释复冰和冰的超润滑现象[65, 66, 69, 70]。认为液态表皮既是冰的润滑剂又是冰的粘结剂。

首先,我们注意到低配位水分子的共价键会自发收缩[54]。由于氧-氧排斥,O:H 非键长度和水分子间距增大并伴随着双重电子极化。低配位氢键的弛豫不仅拓展准固态温区而且在水和冰的表皮,水合层、水滴和汽泡以及水受限在疏水微通道中形成超固态 – 疏水、强极化、高弹性、高粘滞、高熔点、低冰点,低沸点,低密度(≤ 0.75 单位)等[15, 72]。表皮超固态的程度与其曲率正相关。超固态水表皮的 O:H 软声子的高弹性和偶极子的强斥力主导其疏水和冰的超润滑特性[18, 73]。此外,实验已经证明 H—O 键的结合能决定准固态的熔点[17]。受压时,H—O 键伸长弱化,熔点降低,压力撤除后,熔点复原。所以浮冰现象源于氢键的形变自愈合特性。

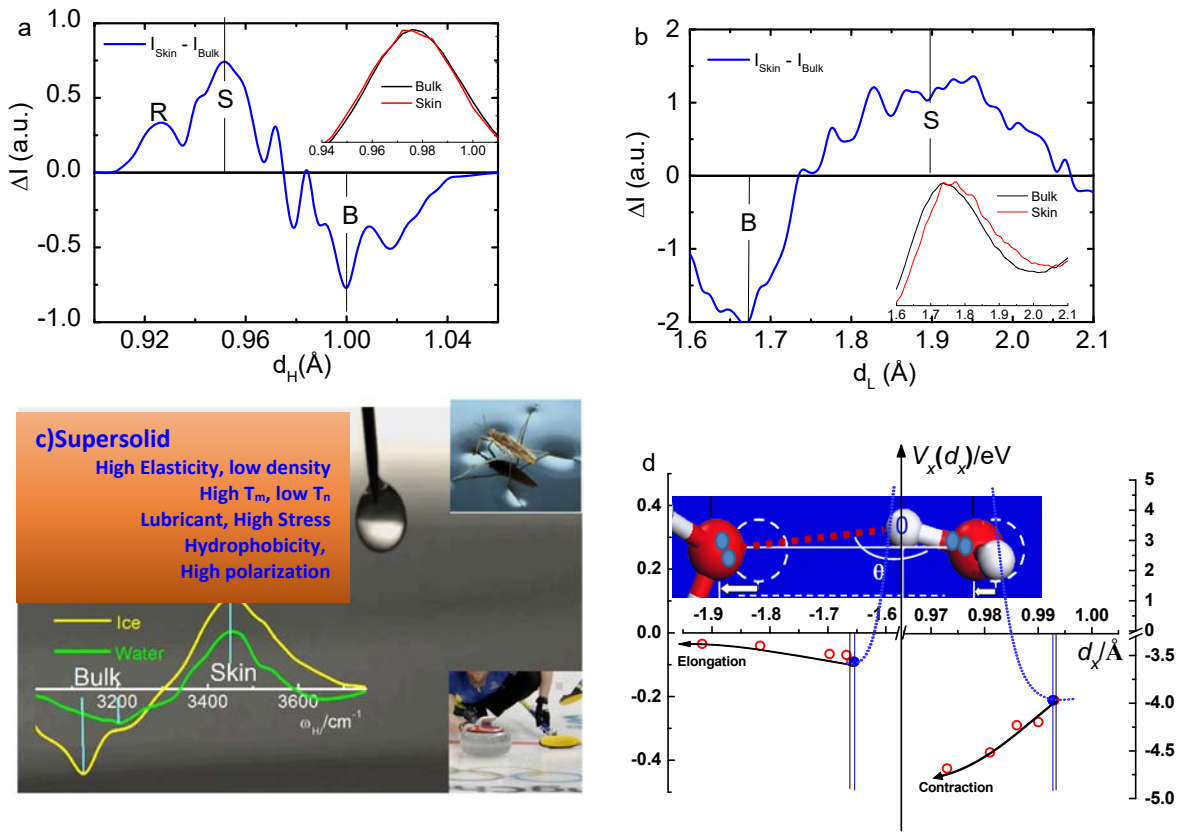

图 4 (a, b)分子动力学计算的 200K 冰表皮 O:H 和 H—O 分段长度相对体相水的协同弛豫[71](B 指体内 S 为表皮 R 为表面悬键), (c) 25℃水和-15℃冰的拉曼差谱[71]揭示分子低配位导致的超固态和 H—O 拉伸共频振动 3450 cm$^{-1}$ 主导冰水的超固态和相应物性。3450 cm$^{-1}$ 峰的面积分代表表皮 O:H—O 键的丰度 – 冰/水表皮厚度比约为 9/4。(d) (H$_2$O)N 团簇分子数目从 6 变到 2 时氢键分段长度和势能路径[45]。

Fig. 4. (a, b) Molecular dynamics derived O:H and H—O segmental lengths for 200 K ice. (c) Raman differential spectra for −15 °C ice and 25 °C water [71] show that ice and water share the same supersolid skin nature characterized by the H—O peak at 3450 cm$^{-1}$. Spectral peak area integration suggests that the the ice/water skin thickness ratio is amounted at 9/4. (d) Lagrangian transformation from the length and frequency to the potential paths for the O:H—O bond under molecular undercoordination from 6 to 2.

图 4a,b 演示分子动力学计算的冰表皮氢键分段长度的协同弛豫,拉曼 H—O 振动差谱,和(H$_2$O)N 团簇分子数目分辨的氢键势能路径。结果表明,冰表皮的 H—O 键从 1.0 收缩到 0.95 埃,O:H 非键从 1.67 膨胀到 1.90 埃并伴有强涨落[71]。 表层 O—O 间距相对体内膨胀 5.9-8.0%[74, 75]。X-光电

子谱测量揭示 H—O 收缩导致氧 1s 芯电子从水的 536.6 深移到表皮水的 538.1 和汽态水的 539.9 电子伏特[15]。氧 1s 芯电子的能移动仅与 H—O 键的结合能变化成正比。同时，非键电子极化使其束缚能从体内的 3.3 降低到表皮的 1.6 电子伏特[76]。图 4c 的拉曼差谱显示 25°C 液态水表皮具有与-15°C 冰表皮共频 − 全同的 H—O 键长度和能量。振动频率分别由体内水的 3200 和体内冰的 3150 蓝移到冰水表皮的共频 3450 $cm^{-1}$ 波数。对差谱的新峰的面积代表所测得的表皮 H—O 键占总测量数目的分数。由积分可以估算水/冰表层厚度比为 4/9。液态水的涨落使得表皮有序层厚度相对冰降低。图 4d 所示为配位数分辨氢键的势能路径也显示水分子配位数目降低使 H—O 发生自发收缩和 O:H 膨胀并伴随结合能的变化。的确，随配位数目减小，H—O 自发收缩刚度增加，O:H 伸长刚度降低。氢键的由于低配位引起的协同弛豫拓展了准固态相边界，使熔点升高冰点降低，也就是通常所说的纳米水滴的过冷和过热现象。

5.3. 压致氢键对称：准固态相边界拓展与复冰

19 世纪 50 年代，法拉第[65]、汤姆逊[77]和福布斯[78, 79] 先后发现了在−10°C 左右，冰受压融化，当压力撤销后将再次结冰的现象："两块冰放在一起，轻轻一压会形成一个整体这种现象对干冰也不会发生。"这种复冰现象是冰固有的性质[66, 77]。自从他们基于经典热力学理论和准液态表皮观点的基础上提出复冰的可能机制以来，人们对于复冰现象依然没有统一的认识[65, 66]。法拉第认为有一层液态水皮包裹着冰。但液态在两冰块之间不能维持而起到焊接剂的作用。

1859 年，法拉第总结了复冰相关的大量实验[65]："应用多种类型的盐进行尝试，将这些盐的常温饱和晶体利用螺旋压力机使之相互挤压，但均未成功。使用的盐类包括：硝酸盐类，铅、氢氧化钾、苏打；硫酸盐类，苏打、氧化镁、铜、锌；明矾；硼砂铵氯化物；碳酸化钾的铁氰化物；苏打碳酸盐；醋酸铅；碳酸化钾和苏打的酒石酸盐等；这些结果只能说明只有水非常特别。复冰的物理成因似乎并非像设计现在这一系列实验时认为的那样。"

1972 年，Holzapfel 首先预测[82]，在压力作用下，H 质子将位于两个 $O^{2-}$ 之间，导致冰 X 从非对称变为对称相。1998 年，Goncharov 等利用原位高压拉曼光谱证实了这一预测[81]。在压强约 60 GPa、温度 100 K 时，冰 VIII 的 O:H—O 键质子实现对称化。O:H 非键 和 H—O 共价键达到等长 (0.11 nm) [52, 83]后，他们继续增大压强，并没有发现声子振频明显偏移。液态水在压强 60 GPa、温度 85 K 时也会发生质子对称化[84]。压强诱导的质子对称化被归因于"质子量子隧道效应"，最终由 $O^{2-}$ 之间的对称双势阱向中心单势阱的转变（图 1）[52, 85-88]。归其原因，受限于 O:H—O 中质子与近临氧原子作用的属性不明， 关注表观多于能量所起的作用 [1]。然而，经历了长达一个半世纪的漫长辩论后，复冰的定量分析依旧未能完成。

利用O:H—O键弛豫原理我们可以给出定量的证据[53, 64]。对于常规物质，压力会使其内含的所有化学键变短，所有的声子都硬化，如碳同素异构体[89]、IV族[90]、III-V族[91]和II-VI族[92]的化合物。

而冰水却不然。压力使弱的O:H非键变短变强、强的 H—O共价键变长变弱，前者缩短幅度远大于后者的伸长，最终质子位于两氧中间[92]，而并非质子的隧穿。

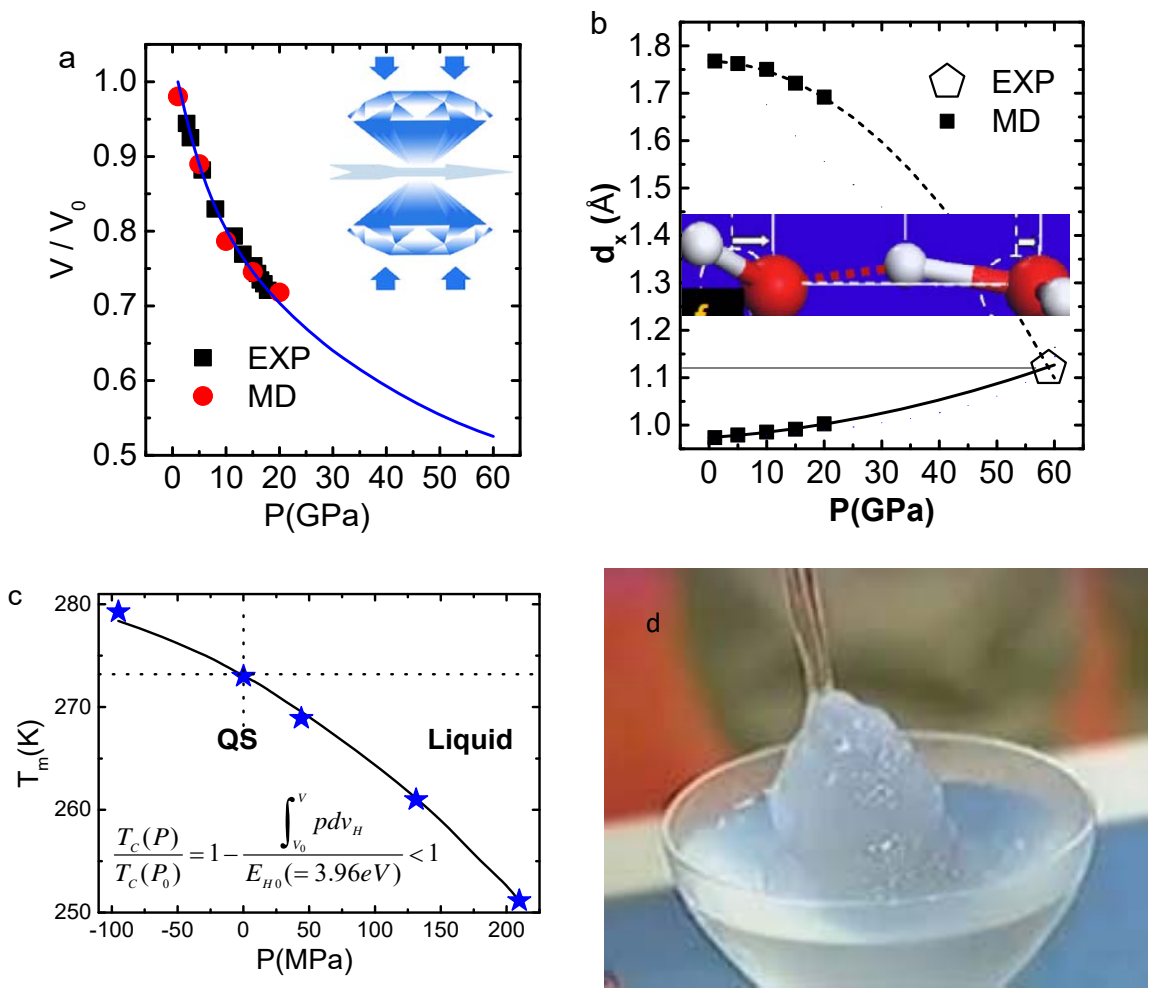

图 5. 冰的(a) 体积和(b) 氢键分段长度的受压弛豫的实验和计算结果[53, 80]。在 59 GPa，质子两侧间距对称 O—O 间距 2.20 Å[52, 81]。(c) 压致准固态/液态相边界拓展（图 2a），降低熔点温度 $T_m$、提高冰点温度 $T_N$，(d) 准固态过冷水受瞬态压强扰动提高 $T_N$ 而结冰。
Fig. 5. Molecular dynamics computation resolved the measured V-P into $d_x$-P showing the H—O elongation and O:h contraction toward phase X with O:H—O symmetrization at 60 GPa. (c) Compression disperses the phase boundary of quasisolid phase, which is responsible for the ice regelation. (d) Freezing of supercooled water under perturbation.

图5a给出了原位高压低温同步辐射XRD和拉曼测量的水(300 K)和冰(77 K)的$V/V_0(P)$ 证据[80]。分子动力学计算将$V/V_0(P)$结果分解成成$d_x/d_{x0}$ - $P$ 关系[92]，如图4b所示。当压强从1 GPa增至20 GPa

时，压力使O:H 非键从1.767 Å缩短至 1.692 Å，同时使H—O 段从0.974 Å 伸长至1.003 Å [93-95]。将$d_x/d_{x0}$ - $P$结果自20 GPa外推，得到氢键双段的交叠点，与实测X相58.6—59.0 GPa时氢键双段长度对称化的位置点精准重叠[52, 80, 81]，此时O—O间距为2.20—2.21 Å [52]。计算与实验结果充分证明，O—O 间的库仑斥力调制着O:H—O键的协同弛豫。质子对称化应为O:H—O键压致非对称弛豫的结果。

液态/准固态的$T_C(P_C)$边界可由下式描述[17]，

$$\begin{cases} \dfrac{\Delta T_C(P)}{T_C(P_0)} = -\sum \dfrac{s_x \int_{P_0}^{P} p \dfrac{\mathrm{d}d_x}{\mathrm{d}p}\mathrm{d}p}{E_{x0}} < 0 \\ \mathrm{d}d_x/\mathrm{d}p > 0 \end{cases}$$

只有$d_H$压致伸长符合这一规则。如图 4c 所示，压致 O:H—O 分段长度对称通过调制各自的德拜温度使准固态相边界拓展 – 熔点降低冰点升高。$T_m(P_C)$曲线的数值重现以及加压 210 MPa 使 $T_m$ 降低−22 °C、−95 MPa 使 $T_m$ 升高 6.5 °C [96, 97]，并视 H 原子直径 0.106 nm 为 H—O 共价键直径[98]，导出 H—O 键能 $E_H$ = 3.97 eV，同时也证实了 H—O 共价键决定液态/准固态转变 的熔点温度$T_m$。O:H 键能决定冰点和沸点而 H—O 键能决定冰点。比如在低气压高山上或在真空中，准固态的熔点升高，冰点和沸点降低，而饱和蒸气压的作用另外计及 [99]。在受到瞬态压强微扰如振动或晃动，准固态水的 $T_N$ 高而结冰。这也说明为什么图 4d 所示的当过冷水从容器中倒出时会结冰。

复冰现象的解析说明水分子的自愈和特性。处于低配位状态时，H—O 共价键自发收缩，O:H 非键伸长长。当两块正在融化的冰相互接触时，两者的表皮水分子易于形成氢键倾向于恢复其原有的四配位状态[54]。一旦在外界刺激下 O:H 非键断开，其 O 原子会寻找到其它原子成键，维持其$sp^3$轨道杂化特性，这与金刚石氧化和金属的氧扩散腐蚀类似，当被氧极化的金属或碳原子被剥时候，氧原子总是寻找性的伙伴重新成键[47]。因此，O:H—O 键不发生任何塑性变形，从受压变形或解离状态自愈合。由于内部电子对的排斥[64]，施压时会提高氢键总能($E_L$+ $E_H$)以存储能量。一旦压力消失或配位恢复，O:H—O 键将弛豫恢复至其最初能态。

5.4. 热水速冷之谜：氢键记忆与水表皮超固态

姆潘巴效应即是在相同实验条件下热水较冷水降温快。亚里斯多德[100]在公元前 350 年首先发现这一现象。美国 Brownridge 博士[101]花了十多年时间做了二十多个实验以寻找控制姆潘巴效应的关键因素。2012 年，英国皇家化学学会组织了一场有两万两千个团队和个人参加的竞赛，并提供一千英镑用以奖励能够合理解释姆潘巴效应的答案。尽管人们提出了许多唯像猜测，虽然过冷和热对流解释占优，但尚无定量证明。

通过求解包含表皮超固态的傅里叶热流体传导方程，我们从理论上再生了实验观测结果。对已知的键长-温度和温度-时间的函数求导得到了键长随时间和初始温度的变化速率。结果表明，姆潘巴现象集合了氢键能量在变温时的"存储-释放-传导-耗散"的系统过程。图 6 所示的理论再生实验结果不仅证明我们对水的温致、压致、和低配位效应处理可靠性而且表明所提出姆潘巴佯谬成立的条件和物理机制成立[102]：

1) 液态水通过 O:H 热膨胀和 H—O 热缩而存储能量
2) 氢键的记忆效应使其能量释放速率与其初始热变形或能量的存储正相关
3) 水的表皮超固态提高热扩散利于热能从体内向外传导
4) 热源向冷库的能量耗散必须经过严格非绝热过程。任何热阻都会妨碍姆潘巴效应发生
5) 其它热对流、蒸发、杂质、过冷、过热等效应的贡献可以忽略。

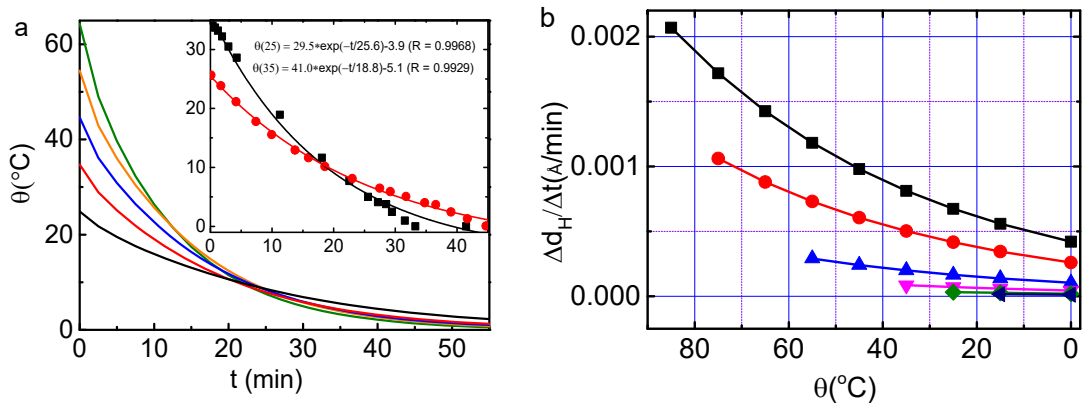

图 6. (a)姆潘巴效应的理论与实验观测(内插)的水的降温曲线随起始温度的变化曲线。交点出现代表此效应发生。(b)氢键长度弛豫线速率与水的初始温度的变化关系[102]。
Fig. 6. Initial temperature dependence of (a) the measured (inset) and calculated temperature as a function of decay time and (b) the linear velocity of the H—O bond as a function of temperature. The cross point in (a) indicates the presence of the Mpamba effect.

6. 结论：启示与展望

基于上述思维方式的探索和实验研究，我们已经获得对于以下列问题为代表的以及酸碱盐水合动力学[103]的自洽定量解。这些进展充分地证明了氢键非对称弛豫必要性和我们所采用的处理方法的合理性。主要进展包括 [1, 15, 17, 18, 99]：

1) 水的双相有序结构 – 超固态表皮包裹的单相四面体配位、高有序、强关联、强涨落可流动"分子单晶"[71]
2) 氢键可近似为超短程，非对称，强关联，氢键振子并显示受激协同弛豫[45, 54]
3) 水与冰的质量密度—几何构型—分子尺度—分子间距的定量关联[104]
4) 氢键-电子-声子-物性的关联与反常物性的起因[15]
5) 相图中各相和相边界处氢键的分段协同弛豫动力学按 T-P 相边界斜率的表述[17]
6) 复冰现象 — 压致熔点漂移 —氢键的超强恢复能力和准固态相边界的漂移[64]
7) 压致氢键的质子对称化 — 外压与 O—O 斥力协同作用下的 O:H 压缩和 H—O 伸长[53]
8) 浮冰现象 – 准固态温区的 H—O 冷缩与 O—O 斥力协同作用下的 O:H 冷胀[55]
9) 质量密度的四温区震荡 — 低比热分段服从热胀冷缩定律，高比热分段反之[55]
10) 低分子配位体系反常热力学（过热与过冷）— H—O 收缩 O:H 膨胀导致的超固态温区弛豫[54, 68, 105]
11) 姆潘巴佯谬 – 氢键记忆热动量，表皮超固态热扩散，非绝热的能量释放-传导-耗散[102]
12) 亲水-疏水受激转变 —水表皮的本征极化以及接触界面偶极子的可控性产生与淹没[73]
13) 超润滑与量子摩擦 — 接触界面偶极子间静电斥力与 O:H 弱声子高振幅低振频超弹性[18]
14) 介电溶质的超固态水合团簇 — 低密度，强极化，高粘度，低涨落，高热稳[59]
15) 准固体相边界可控弛豫 — 氢键分段长度和振动频律决定相应德拜温度[45]
16) 霍夫梅斯特效应 — 盐溶液表面张力与蛋白质溶解能力- 氢键极化与弛豫[106]

水形成如此一个高度有序、强关联、强涨落体系，它不仅包含超短程非对称强耦合作用而且对任何微弱的扰动和辐射都极其敏感。这使它显现多米诺骨牌效应而长程传递信息和形变。事实上，水远比我们所能想象的有趣但是远没有我们想象的那样复杂。作为关键，氢键的协同弛豫与孤对电子极化是如此的简单、重要、无处不在。

虽然我们所做尝试和取得的进展对水的知识海洋贡献微不足道，我们希望她能激发新的思维和处理方法以及更多的兴趣和智慧彻底解决水的所有谜题。继续拓展现有结果和方法去研究水在受不同物质约束，水与软物质，生命体，以及其他物质的作用以及它在生物电子学，医药与食品科学，含能晶体储能燃爆反应，和疾病防护以及对人体的信号加工传递，DNA 修复，离子通道调节以及生物和有机电子学等方面的作用将更加令人兴奋且具有无限的前景。

图文摘要

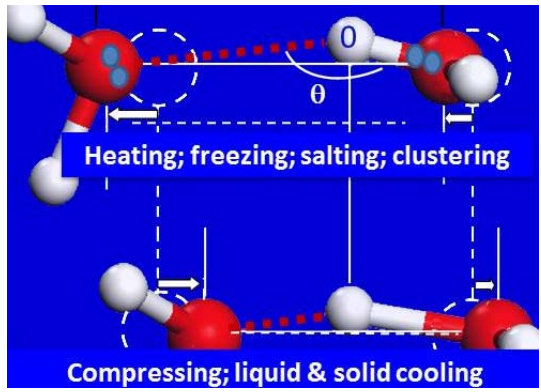

# 参考文献